\documentclass[10pt,a4paper]{article}
\usepackage[utf8]{inputenc}
\usepackage{geometry}
\usepackage{graphicx}
\usepackage{hyperref}
\usepackage{color}

\usepackage{amsmath}
\usepackage{verbatim} 

\begin{document}

\begin{titlepage}
\begin{flushright}
\end{flushright}
\begin{flushright}
\end{flushright}
\vfill
\begin{center}
{\Large\bf Leading Logarithms of the Two Point Function in Massless O(N) and SU(N) Models to any Order from Analyticity and Unitarity}
\vfill
{\bf B. Ananthanarayan$^a$, Shayan Ghosh$^a$, Alexey Vladimirov$^b$, Daniel Wyler$^c$} \\[1cm]
{$^a$ Centre for High Energy Physics, Indian Institute of Science, \\
Bangalore-560012, Karnataka, India}\\[0.5cm]
{$^b$ Institut f\"ur Theoretische Physik, Universit\"at Regensburg,\\
D-93040 Regensburg, Germany}\\[0.5cm]
{$^c$ Physik-Institut, Universit\"at Z\"urich, \\
Winterthurerstrasse 190, \\
CH-8057 Zurich, Switzerland}
\end{center}
\vfill
\begin{abstract}

Leading (large) logarithms in non-renormalizable theories have been investigated in the recent past. Besides some general considerations, explicit results for the expansion coefficients (in terms of leading logarithms) of partial wave amplitudes and of scalar and vector form factors have been given.
Analyticity and unitarity constraints haven been used to obtain the expansion coefficients of partial waves in massless theories, yielding form factors and the scalar two-point function to five-loop order in the O(4)/O(3) model. Later, the all order solutions for the partial waves in any O(N+1)/O(N) model were found. Also, results up to four-loop order exist for massive theories.
Here we extend the implications of analyticity and unitarity constraints on the leading logarithms to arbitrary loop order in massless theories. We explicitly obtain the scalar and vector form factors as well as to the scalar two-point function in any O(N) and SU(N) type models. We present relations between the expansion coefficients of these quantities and those of of the relevant partial waves. Our work offers a consistency check on the published results in the O(N) models for form factors, and
new results for the scalar two-point function.  For the SU(N) type models, we use the known expansion coefficients for partial waves to obtain those for scalar and vector form factors as well as for the scalar two-point function. Our results for the form factor offer a check for the known and future results for massive O(N) and SU(N) type models when the massless limit is taken. Mathematica notebooks which can be used to calculate the expansion coefficients are provided as ancillary files.

\end{abstract}
\vfill
\vfill
\end{titlepage}

\section{Introduction}

Large logarithms are a trademark of radiative loop calculations and often signal substantial modifications of tree level results. It is well known that in renormalizable theories, the so called leading logs (lowest power of the coupling constant) can be conveniently summed up by renormalization group techniques and the result of a simple one-loop calculation.  Indeed, this often leads to large corrections and a substantial reduction of the (artificial) scale dependence. The next-to-leading logarithms can be treated based on two-loop calculations; and  similarly for higher loop orders.

It was pointed out by Weinberg \cite{Weinberg:1966kf} that the coefficients of the leading logarithms in effective (non-renormalizable) field theories at two-loop order could also be obtained from a one-loop calculation. An early application was in demonstrating how to derive these leading logs coefficients from renormalization group (RG) equations for effective field theories by Colangelo \cite{Colangelo:1995np}. Even earlier, Kazakov \cite{Kazakov:1987jp} argued that renormalization group techniques could be used to relate higher order leading logs to a one-loop calculation in a general setting for non-renormalizable
Lagrangians, following notions put forward in 
ref.~\cite{AGFM} in a different context.{\footnotesize }
Soon afterwards,  the double chiral logs of the chiral Lagranigan were evaluated along these lines by  Bijnens, Colangelo and Ecker \cite{Bijnens:1998yu}. 

Encouraged by these developments, Buchler and Colangelo \cite{Buchler:2003vw} extended the consideration of RG invariance to a generic non-renormalizable theory, and  evaluated the leading logarithms for the pion mass and decay constant. An application of these ideas was made in the context of massless O(4)/O(3) theories by Bisseger and Fuhrer\cite{Bissegger:2006ix}, who were able to obtain the leading logarithms in $\pi-\pi$ partial waves, form factors and the two point function of scalar currents to three, four and five loop orders respectively. These authors introduced unitarity and analyticity as powerful tools for extending results at a given order to higher orders; explicitly they gave (chain) relations for iteratively solving for the form factors\footnote{Here and elsewhere this shorthand expression stands for solving for the coefficients of the expansion in leading logarithms.}.

A large step was made by Kivel, Polyakov and Vladimirov \cite{Kivel:2008mf} who demonstrated that RG invariance
allows to determine the leading logarithms of $\pi-\pi$ scattering amplitudes  to arbitrary loop order 
in massless O(N) type model. Consequently, it was shown that in these models one could also obtain expansions for the scalar and vector form factors at any desired order \cite{Kivel:2009az}. 
Subsequently,  the implications of analyticity and unitarity to the pion partial wave amplitudes of definite isospin were studied, and also in arbitrary dimensions \cite{Koschinski:2010mr} \footnote {In fact, unitarity and analyticity imply relations between form factors and partial wave amplitudes; these can be manifest either as i) an Omn\`es type solution for the form factor in terms of the partial wave assuming elastic unitarity, or ii) as a relation between the discontinuity of the form factor and a product of a partial wave with the same form factor.}. Some explorations of massless SU(N) type models (which we use as an abbreviation for $SU(N) \times SU(N)/SU(N)$) have been carried out and the partial waves have been studied in the thesis of Vladimirov \cite{Vladimirov:thesis}; in  \cite{Polyakov:2010pt} application to general $\sigma$-models were  considered.

Massive O(N) and SU(N) theories have been studied by Bijnens and Carloni \cite{Bijnens:2009zi, Bijnens:2010xg},
and Bijnens, Kampf and Lanz \cite{Bijnens:2012hf, Bijnens:2013yca}, with results for both amplitudes and form factors; this method has 
also been generalized to the pion-nucleon system \cite{Bijnens:2014ila}. In the massless limit, the results have been found to agree with the massless results discussed above. In particular, the discontinuity relation (from the analyticity of the amplitude) was used as a consistency check. 

Motivated by all the above, here we extend the results present in the literature for massless O(N) 
models to include the two-point function of scalar currents to arbitrary orders; and give partial waves, form factors and two-point functions for SU(N) models. 
While many results are available, including technical information such as mathematica notebooks to calculate certain quantities, we provide here complete
and explicit codes in order to show the power of the methodology developed and to make these results available for future applications. This is based on the discontinuity equation;
see the previous work by Koschinski, Polyakov and Vladimirov \cite{Koschinski:2010mr} as well as that of Vladimirov \cite{Vladimirov:thesis}). 
Our results provide consistency checks for the results of Bijnens and Carloni when we take the massless limit of their expressions, and can also be used when extensions of their work to higher loops and for the scalar two point function in the O(N) become available available, and for those of Bijnens, Lanz and Kampf for the SU(N) case. Detailed Mathematica notebooks are provided for ease of comparison and reproduction.

The contents of the paper are as follows. In Sec.~2, we set up the notation and define important quantities. In Sec.~3, we  study the discontinuity equations, and provide the recursion relations between the expansion coefficients of the partial waves, the form factors and the two-point functions and solve them subsequently.
In Sec.~4, we consider the O(N) models and apply our relations for the two-point function of the scalar current to present results for general O(N) to a chosen high order up to seven loops, and show that these are consistent with the results of \cite{Bissegger:2006ix} obtained in the O(4)/O(3) model. In Sec.~5, we obtain the scalar and vector form factors in the SU(N) models which have not appeared anywhere in the past. The scalar form factor is then used to obtain the two-point function of scalar currents. This last section requires an input of the partial waves as described in the previous section. We then conclude with a discussion of the results in Sec.~6. In the Appendix, we describe our Mathematica codes, which have been provided as ancillary codes.

\section{Definitions and notation}

The scalar and vector form factors are the matrix elements of the scalar (vector) current between two particles, and are defined as \cite{Bijnens:2013yca}:
\begin{align}
	\langle \phi^a(p_f) | -j_0^s | \phi^a(p_i) \rangle = F_S (s)
\end{align}
and
\begin{align}
	\langle \phi^a(p_f) | j_{V,\mu}^c | \phi^b(p_i) \rangle = \langle T^c (T^b T^a - T^a T^b) \rangle (p_f+p_i)_{\mu} F_V ( s ),
\end{align}
respectively, where $s=(p_f-p_i)^2$, $T^a$ the algebra generators of the particle representation, and $a,b,c$ are the group indices. The leading log expansion for these quantities is given by \cite{Bijnens:2013yca}:
\begin{align}
	F_{S,V}(s) = \sum_{n=0}^{\infty} f_n^{S,V} \hat{S}(s)^n \log^n \left( -\frac{\mu^2}{s} \right) \label{EqSVexp}
\end{align}
where the $f_n^{S,V}$ are coefficients of the leading logs,
$\hat{S}(s) = \frac{s}{(4\pi F)^2}$ is a dimensionless quantity with $F$ being the tree-level pion decay constant.

Unitarity and analyticity dictate that the discontinuity of the scalar and vector form factors satisfy \cite{Bijnens:2012hf, Bissegger:2006ix}:
\begin{align}
	\text{disc } F_S(s) = 2 \; i  \; t_0^0(s) F_S^*(s) \label{EqFSdisc}
\end{align}
and
\begin{align}
	\text{disc } F_V(s) = 2 \; i \; t_1^1(s) F^*_V(s) \label{EqFVdisc}
\end{align}
respectively, where $t_l^I(s)$ is the partial wave amplitude of $\pi\pi$ scattering, and is shown in \cite{Koschinski:2010mr} to be given by:
\begin{align}
	t_l^I(s) = \frac{\pi}{2} \sum_{n=1}^{\infty} \omega_{nl}^{I} \frac{\hat{S}(s)^n}{2l+1} \log^{n-1}\left( \frac{\mu^2}{s} \right) + \mathcal{O}(\text{NLL}) \label{Eqtexp}
\end{align}
where $\omega_{nl}^{I}$ are the leading log coefficients of the pion-pion scattering amplitudes. For definitions of the scattering amplitude for the O(N) model see \cite{Bijnens:2009zi,Bijnens:2010xg}, and for the SU(N) models see \cite{Bijnens:2013yca,Bijnens:2011fm}.

The index $I$ of $t_l^I$ enumerates the irreducible representations of $(R\times R)$, where $R$ is the representation of the scattering particle. In the case of O(N) models, the pions belongs to the defining (or vector) representation and thus the product $(R\times R)$ has three entries. In the terms of iso-spin components of O(4)/O(3) they correspond to isospin of produced system equals to 0,1, and 2, respectively. For SU(N) models the pions belong to adjoint representation, and the product $(R\times R)$ contains 7 irreducible representations, $I=1,..,7$ \cite{Vladimirov:thesis}. 

Aside from the scalar and vector form factors in SU(N), we are also interested in the scalar two point function between two scalar currents in both O(N) and SU(N). In the chiral limit of the $O(4)/O(3)$ model, this is defined by \cite{Bissegger:2006ix}:
\begin{align}
	H(s) = i \int dx \; e^{ipx} \langle 0 | T j_0^s(x) j_0^s(0) | 0 \rangle, \quad s=p^2.
 \end{align}
A leading log expansion can be given for it:
\begin{align}
	 H(s) = \frac{B^2}{16\pi^2} \sum_{n=0}^{\infty} p_n \hat{S}(s)^n \log^n \left( -\frac{\mu^2}{s} \right) \label{EqHexp}
\end{align}
where $p_n$ is the leading log coefficient at order $n$, and where $B$ is related to the quark condensate \cite{Bissegger:2006ix}. In analogy to the expression given in \cite{Bissegger:2006ix} for O(3) and SU(2) the discontinuity across its cut is given by:
\begin{align}
	\text{disc } H(s) = \frac{M i}{16 \pi} \left| F_S(s) \right|^2 \label{EqHdisc}
\end{align}
where the $M$ is an overall multiplicative factor denoting the number of Goldstone bosons arising from the sum over intermediate states. Therefore for O(N+1)/O(N) models, it is given by given by $N$. For the $SU(N) \times SU(N)/SU(N)$ models, $M=N^2-1$.

\section{Relations between expansion coefficients}

In \cite{Kivel:2009az}, building on the method first proposed in \cite{Kivel:2008mf}, expressions for the coefficients of the leading log expansion of the scalar and vector form factors are given as recursion relations that depend on the partial wave decomposition of the crossing matrix $\Omega_p^{l i}$ (see Eq.(\ref{EqJETPfs}) below for instance). By developing the discussion given in \cite{Bissegger:2006ix}, these recursion relations are solved to provide expressions for $f_n^{S,V}$ in terms of the leading log coefficients of the $\pi-\pi$ scattering amplitudes $\omega_{nl}^{I}$. The solution provided is an integral equation of Omn\`es type, which although formally correct, is not particularly useful for practical applications. However, the same unitarity relation used to derive the Omn\`es type solution can be expressed in terms of the discontinuities stated earlier, Eq.(\ref{EqFSdisc}) and (\ref{EqFVdisc}). By explicitly introducing the leading log expansions for the form factors Eq.(\ref{EqSVexp}) on the LHS of the discontinuity equations and isolating the leading log at a given order, and on the RHS introducing the expansions of the partial waves Eq.(\ref{Eqtexp}) and the form factors themselves, it is possible to extract recursion relations for $f_n^{S,V}$ that depends only on the $\omega_{nl}^{I}$. This has been done in \cite{Vladimirov:thesis}, and we have reproduced it as well.

For the scalar form factor, this solution reads:
\begin{align}
	f_n^S(N) = \frac{1}{2(n-1)} \sum_{i=1}^{n-1} \omega_{i0}^{I=0} f_{n-i}^S(N) \label{Eqfs}
\end{align}
and for the vector form factor, it reads:
\begin{align}
	f_n^V(N) = \frac{1}{6(n-1)} \sum_{i=1}^{n-1} \omega_{i0}^{I=1} f_{n-i}^V(N) \label{Eqfv}
\end{align}
where $f_1^S(N) =f_1^V(N) =1$, and $n$ denotes the loop order.

Using a similar approach, we obtain for the leading log coefficients of the scalar correlator, the following result:
\begin{align}
	p_n(N)= \frac{M}{2(n-1)} \sum_{i=1}^{n-1} f_i^S  f_{n-i}^S \label{Eqpn}
\end{align}
Eq.(\ref{Eqpn}) is one of the principal results of this paper, and has not appeared
before in the literature.  

\section{O(N)}

In \cite{Kivel:2009az}, the following recursion relation was used to generate the coefficients of the form factor expansion:
\begin{align}
	f_n^{S,V} = \frac{1}{n} \sum_{m=0}^{n-1} \sum_{\substack{j=0 \\ j \text{ even}}}^{n-m} \Gamma_{S,V}^{(n-m,j)} f_{m}^{S,V} \omega_{(n-m-1)j} \label{EqJETPfs}
\end{align}
where
\begin{align}
	\Gamma_S^{(p,l)} = \frac{N}{2} \delta_{l0} + \Omega_p^{l0}
\end{align}
or
\begin{align}
	\Gamma_V^{(p,l)} = \frac{1}{3} \Omega_p^{l1}.
\end{align}
Here, the coefficient $\omega_{nl}=(\omega_{nl}^0-\omega_{nl}^2)/N$. The functions $\Omega_{p}^{ll'}$ are the components of the crossing matrix $(s\leftrightarrow t)$ for the amplitude in the partial wave basis. The values for coefficients $f_n^{S,V}$ are given in Table I of \cite{Kivel:2009az} for the scalar form factor. The values of this table were reproduced by solving Eq.(\ref{Eqfs}) with its initial condition $f_1^{S}=1$. Due to an extra $(-1)^{p+1}$ factor in the the expression for $\Gamma_V^{(p,l)}$ in \cite{Kivel:2009az}, Table II of the same paper (which tabulates $f_n^V$) is erroneous. Solving the recursion relation of Eq.(\ref{Eqfv}) produces the same set of values of the expansion coefficients as using the correct $\Gamma_V^{(p,l)}$ in Eq.(\ref{EqJETPfs}) does.

Using the values of $f_n^S$ generated from Eq.(\ref{Eqfs}), and Eq.(\ref{Eqpn}), we obtain the coefficients of the leading log expansion of the scalar current $H(s)$. We tabulate these in Table~\ref{TableONp} as they have hitherto not appeared in the literature. We provide values of $p_n$ for both general O(N+1)/O(N) and for O(4)/O(3), upto 7-loop order. These values are the generalization to arbitrary loop order $n$, and arbitrary group dimension $N$, of the results of \cite{Bissegger:2006ix}. 

We provide a supplementary Mathematica notebook \texttt{ON.nb} in which the results of this section are demonstrated. In this notebook, we provide a solution of the recursion relations Eq.(\ref{Eqfs}) and Eq.(\ref{Eqfv}), which can be used to generate solutions to any chosen order $n$. In the same notebook, we also encode the complete list of partial wave expansion coefficients which can be generated based on the expressions give in \cite{Koschinski:2010mr} for d=4. These are then used as input to generate the expansion coefficients for the scalar and vector form factors. We have a separate module implementing the effective Lagrangian expressions given in \cite{Kivel:2009az} that do not rely on partial wave amplitudes. The results are mutually consistent. Having thus obtained the expansion coefficients of the scalar form factor, they are in turn used to produce the expansion coefficients of the two point function of scalar currents, $p_n$. This represents the full generalization of the 5-loop results of Bijnens and Carloni to arbitrary N and an arbitrary number of loops.

Up to and including 4 loops, these agree with the massless limit of the expressions of Bijnens and Carloni. By exhibiting the results to higher loops, we provide a consistency check for the massive results, should they be worked out to yet higher loops.  Furthermore, the scalar two-point function has not been computed in the massive case, and our results can be used as consistency check at any order of the massless limit of a massive computation for this quantity should it be done.

\begin{table}
\center
\renewcommand{\arraystretch}{1.5}
\begin{tabular}{|c|c|c|}
\hline 
$n$ &  $p_n(N)$ & $p_n(3)$ \\[1ex]
\hline
\hline
1 & 0 & $0$ \\ 
\hline 
2 & $2 M$ & $6$ \\ 
\hline 
3 & $M N-M$ & $6$ \\ 
\hline 
4 & $\frac{M N^2}{2}-\frac{47 M N}{54}+\frac{10 M}{27}$ & $\frac{61}{9}$ \\ 
\hline 
5 & $\frac{M N^3}{4}-\frac{553 M N^2}{864}+\frac{41 M N}{72}-\frac{155 M}{864}$ & $\frac{68}{9}$ \\ 
\hline 
6 & $\frac{M N^4}{8}-\frac{22537 M N^3}{54000}+\frac{1080211 M N^2}{1944000}-\frac{171647 M N}{486000}+\frac{174709 M}{1944000}$ & $\frac{140347}{16200}$ \\ 
\hline 
7 & $\frac{M N^5}{16}-\frac{55883 M N^4}{216000}+\frac{53160113 M N^3}{116640000}-\frac{149841121 M N^2}{349920000}+\frac{75199037 M N}{349920000}-\frac{3235559 M}{69984000}$ & $\frac{897079}{91125}$ \\
\hline
\end{tabular}
\caption{Expansion coefficients $p_n(N)$ Eq.(\ref{Eqpn} of 2-point scalar current $p_n$  in $O(N+1)/O(N)$ models for general $N$, with $M=N$, and for $N=M=3$, up to n = 7 loops.}
\label{TableONp}
\end{table}

\section{SU(N)}

Since SU(N) models are considerably less studied than their O(N) counterparts, we recall a few salient features of them. In particular, the two-particle scattering amplitude is given by two functions of the Mandelstam variables which parameterize independent group structures. The leading log coefficients for these functions are denoted in the following as $\omega_{nl}$ and $v_{nl}$. They could be decomposed over seven irreducible representations, $\omega_{nl}^I$ \cite{Vladimirov:thesis}. Of these, two will play a special role for us, namely those corresponding to I=1 and 5. These are such that their S and P waves  enter the discontinuity equations for the scalar and vector form factors, respectively.

For the case of $N=2$, the seven amplitudes collapse to three and have a direct correspondence with those of the $O(N)$ model. In the massless theory, the amplitudes that correspond to B and C are the charges denoted by $\omega$ and $v$. Recall that in the O(N) model there is a single change and it has a closed evolution equation.  In contrast, the  charges in the SU(N) theory have a coupled system of closed evolution equations, and in \cite{Vladimirov:thesis} these are provided.  Furthermore, in the large $N$ limit there is an essential simplification. From the evolution
equations for arbitrary $N$ for the $\omega$ and $v$, one can obtain corresponding solutions for the $\omega^I$ which wires in the full crossing matrices for the seven amplitudes, as well as for all the waves that enter a particular order. It is this system that is solved and used by us in our computations, both for the $\omega$, which are then inserted into the relations for the expansion coefficients for the scalar and vector form factors and eventually for the two-point function. The equations for the charges can be used directly in the direct formula Eq.(\ref{EqBM}) to obtain the vector form factor in general as well as in the large N limit.

An expression for the vector form factor in SU(N), analogous to Eq.(\ref{EqJETPfs}) and which is based on a recursion relation is
\begin{align}
	\Gamma_V^{(p,l)} = \frac{1}{3} \Omega_p^{l1},
\end{align}
\begin{align}
	f_n^V = \frac{1}{n} \sum_{m=0}^{n-1} \sum_{\substack{j=0 \\ j \text{ even}}}^{n-m}  \Gamma_V^{(n-m,j)} f_{m+1}^V \left( \frac{N}{2} \omega_{n-m-1,j} + 2 v_{n-m-1,j} \right) \label{EqBM}
\end{align}
where $f_0^V = 1$ and $n = 0,1,2,3...$. We see that this recursion formula generates the entries given in Table~\ref{TableSUNfsfv} for $f_n^V$.


As described in \cite{Vladimirov:thesis}, in the large-N limit, the full set of coupled RG equations is replaced by a simpler one, $\omega_{nl}^{\text{large-}N} $. This leads to a simpler set of equations for $f^V_n$ which produce the correct coefficient for the highest power of N as found from Eq.(\ref{EqBM}). For completeness, Eq.(\ref{EqBM}) is replaced by the following:
\begin{align}
	f_n^V {}^{\text{large-}N} = \frac{1}{n} \sum_{m=0}^{n-1} \sum_{\substack{j=0 \\ j \text{ even}}}^{n-m}  \Gamma_V^{(n-m,j)} f_{m+1}^V \frac{N}{2} \omega_{n-m-1,j}^{\text{large-}N}. \label{EqLargeN}
\end{align}
and shown in our supplementary notebook \texttt{SUN.nb} to agree with expected values.
This result is obtained by using the $\beta$-functions presented in the thesis of
Vladimirov~\cite{Vladimirov:thesis} and is obtained without the use of unitarity
and analyticity.

The solutions for the $\omega$ and $v$, for use in Eq.(\ref{EqBM}) as well as in Eqs.(\ref{Eqfs}) and (\ref{Eqfv}), are obtained by solving the RG equations, and inserting into the partial wave expansion coefficients of the seven irreducible amplitudes following the expressions given in \cite{Vladimirov:thesis}. The corresponding coefficients are $\omega_i$, $i=1,...,7$. Notice that in Eq.(III-2-51) of \cite{Vladimirov:thesis}, the $1/d_I$ factor should not be present, and that the inverse relation for $v_n$ should have a factor of 1/4.

We also note that for the unitarity relation, we define $\omega^S = \omega_1$ for the scalar, and for the vector the corresponding wave will be denoted by $\omega^V = \omega_5$. With these choices, we can now use Eqs.(\ref{Eqfs})-(\ref{Eqfv}). In Tables~\ref{TableSwave} and~\ref{TablePwave}, we provide a table upto 7-loop order for the S-wave of the scalar and the P-wave of the vector, respectively, for both general SU(N) and SU(2) models. We now insert these quantities into the unitarity relations, solve them, and tabulate the scalar and vector form factors to the same order. These are listed in Table~\ref{TableSUNfsfv}. In the Mathematica notebooks we provide, we perform consistency checks by putting N=2 and compare them with the corresponding results for the orthogonal models with N=3. In \cite{Bijnens:2013yca} are listed $f^S_n$ and $f^V_n$ for massive SU(N) models to 5-loop order. By taking their massless limit and compare against the entries of Table~\ref{TableSUNfsfv}, we find perfect agreement. Finally, in Table~\ref{TableSUNp}, we list the expansion coefficients of the two-point scalar currents for SU(N) and SU(2) models. 

\begin{table}
\center
\renewcommand{\arraystretch}{1.5}
\begin{tabular}{|c|c|c|} 
\hline 
$n$ &  $SU(N)$ & $SU(2)$ \\[1ex]
\hline
\hline
1 & $N$ & $2$ \\ 
\hline 
2 & $\frac{25}{36} N^2$ & $\frac{25}{9}$ \\ 
\hline 
3 & $\frac{25}{72} N^3$ & $\frac{25}{9}$ \\ 
\hline 
4 & $\frac{155887}{777600} N^4+\frac{1129}{7200} N^2$ & $\frac{18637}{4860}$ \\ 
\hline 
5 & $\frac{5130707}{46656000} N^5+\frac{30707}{1296000} N^3$ & $\frac{540707}{145800}$ \\ 
\hline 
6 & $\frac{5675249833}{91445760000} N^6+\frac{255657313}{5080320000} N^4+\frac{3515233}{60480000} N^2$ & $\frac{357894863}{71442000}$ \\ 
\hline 
7 & $\frac{60834697957}{1755758592000} N^7+\frac{34139562617}{2560481280000} N^5+\frac{1379751769}{142248960000} N^3$ & $\frac{59282627549}{12002256000}$ \\
\hline
\end{tabular}
\caption{S-wave ($\omega^S$) for scalars in SU(N) and SU(2) models. These have been calculated in the Mathematica notebook \texttt{SUN.nb}.}
\label{TableSwave}
\end{table}

\begin{table}
\center
\renewcommand{\arraystretch}{1.5}
\begin{tabular}{|c|c|c|} 
\hline 
$n$ &  $SU(N)$ & $SU(2)$ \\[1ex]
\hline
\hline
1 & $\frac{N}{2}$ & 1 \\ \hline 
2 & 0 & 0 \\ \hline
3 & $\frac{1}{64}N^3+\frac{9}{16} N$ & $\frac{5}{4}$ \\ \hline
4 & $-\frac{901}{518400} N^4 - \frac{901}{14400} N^2 $ & $-\frac{901}{3240}$ \\ \hline 
5 & $\frac{291143}{435456000} N^5 + \frac{752987}{8064000} N^3 + \frac{871}{3840} N $ & $\frac{207871}{170100}$ \\ \hline 
6 & $-\frac{495203}{4354560000} N^6 - \frac{8733881}{282240000} N^4 + \frac{61529}{11760000} N^2$ & $-\frac{22931351}{47628000}$  \\ \hline
7 & $\frac{187257853}{5852528640000} N^7 + \frac{6816041549}{320060160000} N^5 + \frac{77753874571}{1706987520000}  N^3 + \frac{604493021}{6773760000} N $ & $\frac{4914744211}{4000752000}$ \\ \hline
\end{tabular}
\caption{P-wave ($\omega^V$) for vectors in SU(N) and SU(2) models. These have been calculated in the Mathematica notebook \texttt{SUN.nb}.}
\label{TablePwave}
\end{table}

\begin{table}
\center
\renewcommand{\arraystretch}{1.5}
\begin{tabular}{|c|c|c|}
\hline 
$n$ &  $f_n^S$ & $f_n^V$ \\[1ex]
\hline
\hline
1 & 1 & 1 \\ 
\hline 
2 & $\frac{1}{2} N$ & $\frac{1}{12} N$ \\ 
\hline 
3 & $\frac{43}{144} N^2$ & $\frac{1}{288} N^2$ \\ 
\hline 
4 & $\frac{143}{864} N^3$ & $\frac{5}{5184} N^3 + \frac{1}{32} N$ \\ 
\hline 
5 & $\frac{580837}{6220800} N^4 + \frac{1129}{57600} N^2$ & $\frac{1}{518400} N^4 - \frac{1}{345600} N^2$ \\ 
\hline 
6 & $\frac{48727189}{933120000} N^5 + \frac{315439}{25920000} N^3$ & $\frac{126059}{6531840000} N^5 + \frac{545009}{181440000} N^3 + \frac{871}{115200} N$ \\ 
\hline 
7 & $\frac{1606873337}{54867456000} N^6 + \frac{684700627}{60963840000} N^4 + \frac{3515233}{725760000} N^2$ & $-\frac{102731}{94058496000} N^6 - \frac{70609843}{121927680000} N^4 + \frac{39629}{31360000} N^2$ \\ 
\hline 
\end{tabular}
\caption{$f^S$ (Eq.(\ref{Eqfs})) and $f^V$ (Eq.(\ref{Eqfv})) for SU(N) models.}
\label{TableSUNfsfv}
\end{table}

\begin{table}
\center
\renewcommand{\arraystretch}{1.5}
\begin{tabular}{|c|c|c|}
\hline 
$n$ &  $p_n(N)$ & $p_n(2)$ \\[1ex]
\hline
\hline 
1 & $0$ & $0$ \\ 
\hline 
2 & $2 M$ & $6$ \\ 
\hline 
3 & $M N$ & $6$ \\ 
\hline 
4 & $\frac{61 M}{108} N^2$ &  $\frac{61}{9}$ \\ 
\hline 
5 & $\frac{17 M}{54} N^3$ & $\frac{68}{9}$ \\ 
\hline 
6 & $\frac{1372987 M}{7776000} N^4 + \frac{1129 M}{72000} N^2$ & $\frac{140347}{16200}$ \\ 
\hline
7 & $\frac{17300933 M}{174960000} N^5 + \frac{71183 M}{4860000} N^3 $ & $\frac{897079}{91125}$ \\
\hline
\end{tabular}
\caption{Expansion coefficients $p_n(N)$ Eq.(\ref{Eqpn}) of 2-point scalar current in $SU(N)$ models for general $N$ and $M=N^2-1$, and $N=2,\, M=3$, up to $n = 7$ loops.}
\label{TableSUNp}
\end{table}

Analogous to the comments we have provided for the O(N) case, the results in the U(N) case offer checks to the published results
of Bijnens, Kampf and Lanz and also provide a basis for comparison with a future computation to higher loops or for the scalar two-point function.

\section{Discussion and Conclusions}

In this paper we have revisited the issue of leading logarithms in non-renormalizable theories, motivated first by the work of Buchler and Colangelo \cite{Buchler:2003vw} who had discussed the issue in general, and which was subsequently applied in massive O(N) and SU(N) theories by Bijnens, Carloni, Kampf and Lanz \cite{Bijnens:2009zi,Bijnens:2010xg,Bijnens:2012hf,Bijnens:2013yca}. These considerations were themselves motivated by the observation of Weinberg, who showed that two-loop leading logarithms are governed by one-loop quantities, and a formal discussion is given by Kazakov \cite{Kazakov:1987jp}, which was worked out in \cite{Buchler:2003vw}. In \cite{Bissegger:2006ix}, Bissegger and Fuhrer introduced the powerful tools of analyticity and unitarity to obtain results up to 5-loop order in O(4)/O(3) theories. Then, Kivel, Polyakov and Vladimirov \cite{Kivel:2008mf} showed that in massless theories it is even possible to solve for the expansion coefficients of the scattering amplitude to all orders, and applied this to scalar and vector form factors in O(N) models in \cite{Kivel:2009az}. Using analyticity and unitarity, Koschinski, Polyakov and Vladimirov worked out the expansion coefficients of the isospin-like amplitudes \cite{Koschinski:2010mr}. Partial results for the two types of scattering amplitude functions discussed in Section 5 (for SU(N)) were presented in the doctoral thesis of Vladimirov \cite{Vladimirov:thesis}. 

In this paper, that can be seen as a completion of  previous work, we have done the following. We started out by applying analyticity and unitarity constraints which give a relation for the discontinuity of the form factors Eq.(\ref{EqHdisc}), to inter-relate the expansion coefficients of the form factors with those of the partial waves.  We have applied the relations to the scalar and vector form factors of O(N) models and have reproduced the published results of \cite{Kivel:2009az}.  However, this set of relations has not been published before. They are useful in themselves as they offer an important consistency check on the results. The analogous relationship for the 2-point function, Eq.(\ref{Eqpn}), gives the expansion coefficients in terms of those of the scalar form factor.  We have used this to generalize the results in \cite{Bissegger:2006ix} to all orders, and to all O(N) models. 

We then took up SU(N) models for which there are few explicit results in the literature. Using the known relations between the expansion coefficients of the isospin-like S and P waves and the relations Eq.(\ref{Eqfs}), Eq.(\ref{Eqfv}) we obtain the scalar and vector form factors; the expansion coefficients are tabulated extensively. The results agree with those of Bijnens, Kampf and Lanz \cite{Bijnens:2013yca} in the massless limit. We then solve equation Eq.(\ref{Eqpn}) for the 2-point function and provide the results for the expansion coefficients in table 5. In the limit of SU(2)$\times$SU(2) all our results agree with those of O(4)/O(3), for which results exist in the literature although often to limited loop order. We have also found a formula for the vector form factor in analogy with those of \cite{Kivel:2009az}, which also admits a large N version and is consistent with the full results obtained from the solutions of the discontinuity relation.

Indeed, it has been pointed out by Bissegger and Fuhrer that the discontinuity equation can be applied in the following chain of expansion coefficients of increasing loop order to find the leading logarithms in massless $O(4)/O(3)$ (albeit valid all O(N) and SU(N) models) $$t^0_0 \stackrel{\rm disc} \rightarrow  F_S  \stackrel{\rm disc} \rightarrow H, $$ and was applied when the first was known to 3 loops, and therefore yielding the others to 4 and 5 loops, respectively. Now, that the former is now known to all orders, the repeated application of this can yields all order results in O(N) as well as in SU(N) models. For vector form factors, the analogous chain stops after  $t^1_1$ and $F_V$ since there is no vector counterpart for $H$.

We have also provided systematic Mathematica notebooks that can be used to obtain the quantities of interest to any loop order. As mentioned in the introduction, the Mathematica notebooks we provide can allow one to generate solutions to any order for the partial waves, the form factors and the scalar two point function, and can be used to predict the massless limit of the massive computations done in the past for these quantities, and also for the scalar two point function, should they be carried out. This simplification has been traced to the vanishing of tadpoles in massless theories. To this extent, the continuing reason for coming back to the topic of large logarithms in non-renormalizable theories is to gain further insights into this intriguing issue. There is substantial interest in such theories when going beyond the standard model and results are welcome. The scattering amplitude being the building block of the loop expansion, therefore, provides the analog of the coupling constant, while the scalar and vector form factors expansion coefficients enjoy a status similar to those of the expansion coefficients of the anomalous dimensions.

Thus, we see that in the context of effective theories, complete results for the leading logarithms are obtained, with an essential simplification arising in massless theories. Our results indicate that there is a striking similarity to results in renormalizable theories where a simple factorization of the leading logarithm contributions occurs. Here, factorization seems to stem from unitarity and analyticity, which are general properties of field theories. Also, simplifications occur for large N, possibly allowing to study properties of field theories at large N.

As a possible future extension, it would be worthwhile to study how the cancellation of non-local divergences also provide the same kind of factorization properties. Similarly, the solutions found here may be used to further the study of infrared logarithms of sigma models on Riemannian manifolds \cite{Polyakov:2010pt}; other related effects may occur in other important effective theories, such as gravity \cite{Donoghue:2017ovt, Codello:2015pga}.

\section*{Acknowlegements}

We thank Johan Bijnens and Gilberto Colangelo for discussions and correspondence on the subject.  
BA thanks Sachindeo Vaidya for discussions during an early stage of this investigation.
BA thanks the University of Zurich under the Pauli Centre Visitor programme and DW thanks the Indian Institute of Science. Bangalore for hospitality. BA is partly supported by the MSIL Chair of the Division of Physical and Mathematical Sciences, Indian Institute of Science.
\appendix 

\section{Description of Notebooks}

We provide along with this paper two Mathematica notebooks that can be used to generate the results given in the various tables of this paper, but also to higher loop order.  These have
been lodged as ancillary files along with the arxiv submission.

In the notebook \texttt{ON.nb}, building on the notebook provided with the paper \cite{Kivel:2009az}, several formula pertaining to the O(N) models are coded in. In the section ``Expansion coefficients of $f^S$ and $f^V$ using recursion relations of JETP paper", the expression from \cite{Kivel:2009az} are coded in to generate the scalar form factors, and the expression from \cite{Bijnens:2013yca} to generate the vector form factors. The section ``Reproduction of $\omega^I$ using unitarity, analyticity and crossing" has the appropriate formulas of \cite{Koschinski:2010mr} entered and Tables I, II and III of the aforementioned paper reproduced. The next two sections, ``The expansion coefficients of $f^S$ and $f^V$ using recursion relations of the Bissegger \& Fuhrer paper" and ``Expansion coefficients of 2-point scalar currents" encode Eqs.(\ref{Eqfs})-(\ref{Eqpn}) of this paper, generate values of $f^S_n$, $f^V_n$ and $p_n$, and for the first two show that these are equal to the values obtained in the earlier sections.

In the notebook \texttt{SUN.nb}, several formula pertaining to the SU(N) models are coded in. In the section ``Generating the $\omega_1$ and $\omega_5$, the notebook provided with the paper \cite{Kivel:2009az} is built upon by adding expressions from \cite{Vladimirov:thesis} to produce the S-wave and P-wave components of $\Omega$ for the SU(N). In the sections ``Solving the recursion relation" and ``Expansion coefficients of 2-point scalar currents", Eqs.(\ref{Eqfs})-(\ref{Eqpn}) are entered and the values produced using them are shown to be equivalent to the ones in the literature. The last two sections use Eqs.(\ref{EqBM}) and (\ref{EqLargeN}) to reproduce $f_n^V$ and its large N approximation and show that they agree with the results previously obtained.


\newpage

\begin{thebibliography}{99}

\bibitem{Weinberg:1966kf}
  S.~Weinberg,
  Phys.\ Rev.\ Lett.\  {\bf 17} (1966) 616.
  doi:10.1103/PhysRevLett.17.616

\bibitem{Colangelo:1995np}
  G.~Colangelo,
  Phys.\ Lett.\ B {\bf 350} (1995) 85
   Erratum: [Phys.\ Lett.\ B {\bf 361} (1995) 234]
  doi:10.1016/0370-2693(95)00349-P, 10.1016/0370-2693(95)01162-J
  [hep-ph/9502285].

\bibitem{Kazakov:1987jp}
  D.~I.~Kazakov,
  Theor.\ Math.\ Phys.\  {\bf 75} (1988) 440
   [Teor.\ Mat.\ Fiz.\  {\bf 75} (1988) 157].
  doi:10.1007/BF01017179
 
\bibitem{AGFM}
  L.~Alvarez-Gaume, D.~Z.~Freedman and S.~Mukhi,
  Annals Phys.\  {\bf 134} (1981) 85.
  doi:10.1016/0003-4916(81)90006-3
    
\bibitem{Bijnens:1998yu}
  J.~Bijnens, G.~Colangelo and G.~Ecker,
  Phys.\ Lett.\ B {\bf 441} (1998) 437
  doi:10.1016/S0370-2693(98)01193-9
  [hep-ph/9808421].


\bibitem{Buchler:2003vw}
  M.~Buchler and G.~Colangelo,
  Eur.\ Phys.\ J.\ C {\bf 32} (2003) 427
  doi:10.1140/epjc/s2003-01390-2
  [hep-ph/0309049].

\bibitem{Bissegger:2006ix}
  M.~Bissegger and A.~Fuhrer,
  Phys.\ Lett.\ B {\bf 646} (2007) 72
  doi:10.1016/j.physletb.2007.01.025
  [hep-ph/0612096].

\bibitem{Kivel:2008mf}
  N.~Kivel, M.~V.~Polyakov and A.~Vladimirov,
  Phys.\ Rev.\ Lett.\  {\bf 101} (2008) 262001
  doi:10.1103/PhysRevLett.101.262001
  [arXiv:0809.3236 [hep-ph]].

\bibitem{Kivel:2009az}
  N.~A.~Kivel, M.~V.~Polyakov and A.~A.~Vladimirov,
  JETP Lett.\  {\bf 89} (2009) 529
  doi:10.1134/S0021364009110022
  [arXiv:0904.3008 [hep-ph]].

\bibitem{Koschinski:2010mr}
  J.~Koschinski, M.~V.~Polyakov and A.~A.~Vladimirov,
  Phys.\ Rev.\ D {\bf 82} (2010) 014014
  doi:10.1103/PhysRevD.82.014014
  [arXiv:1004.2197 [hep-ph]].

\bibitem{Vladimirov:thesis}
A.A. Vladimirov,``Infrared Lagarithms in Effective Field Theoriesâ",
PhD Theses, Ruhr-University Bochum (2010)

\bibitem{Polyakov:2010pt}
  M.~V.~Polyakov and A.~A.~Vladimirov,
  Theor.\ Math.\ Phys.\  {\bf 169} (2011) 1499
  doi:10.1007/s11232-011-0126-7
  [arXiv:1012.4205 [hep-th]].
  
\bibitem{Bijnens:2009zi}
  J.~Bijnens and L.~Carloni,
  Nucl.\ Phys.\ B {\bf 827} (2010) 237
  doi:10.1016/j.nuclphysb.2009.10.028
  [arXiv:0909.5086 [hep-ph]].
  
\bibitem{Bijnens:2010xg}
  J.~Bijnens and L.~Carloni,
  Nucl.\ Phys.\ B {\bf 843} (2011) 55
  doi:10.1016/j.nuclphysb.2010.09.019
  [arXiv:1008.3499 [hep-ph]].
  
\bibitem{Bijnens:2012hf}
  J.~Bijnens, K.~Kampf and S.~Lanz,
  Nucl.\ Phys.\ B {\bf 860} (2012) 245
  doi:10.1016/j.nuclphysb.2012.02.014
  [arXiv:1201.2608 [hep-ph]].
  
\bibitem{Bijnens:2013yca}
  J.~Bijnens, K.~Kampf and S.~Lanz,
  Nucl.\ Phys.\ B {\bf 873} (2013) 137
  doi:10.1016/j.nuclphysb.2013.04.012
  [arXiv:1303.3125 [hep-ph]].

\bibitem{Bijnens:2014ila}
  J.~Bijnens and A.~A.~Vladimirov,
  Nucl.\ Phys.\ B {\bf 891} (2015) 700
  doi:10.1016/j.nuclphysb.2014.12.015
  [arXiv:1409.6127 [hep-ph]].
    
\bibitem{Bijnens:2011fm}
  J.~Bijnens and J.~Lu,
  JHEP {\bf 1103} (2011) 028
  doi:10.1007/JHEP03(2011)028
  [arXiv:1102.0172 [hep-ph]].

\bibitem{Donoghue:2017ovt}
  J.~Donoghue,
  Scholarpedia {\bf 12} (2017) no.4,  32997.
  doi:10.4249/scholarpedia.32997


\bibitem{Codello:2015pga}
  A.~Codello and R.~K.~Jain,
  Class.\ Quant.\ Grav.\  {\bf 34} (2017) no.3,  035015
  doi:10.1088/1361-6382/aa549d
  [arXiv:1507.07829 [astro-ph.CO]].

\end{thebibliography}
\end{document}